\documentclass[showpacs,preprintnumbers,twocolumn,amsmath,amssymb]{revtex4-1}
\usepackage{graphicx}
\usepackage{dcolumn}
\usepackage{bm}

\begin{document}

\title{Polarized-Deuteron Stripping Reaction at Intermediate Energies}
\author{Valery I. Kovalchuk}
\affiliation{Department of Physics, Taras Shevchenko National University, Kiev 01033, Ukraine}


\begin{abstract}
A general analytical expressions for the cross-section and the polarization of nucleons arising in
the inclusive deuteron stripping reaction have been derived in the diffraction approximation.
The nucleon-nucleus phases were calculated in the framework of Glauber formalism and making use of
the double-folding potential. The tabulated distributions of the target nucleus density and the
realistic deuteron wave function with correct asymptotic at large nucleon-nucleon distances were used.
The calculated angular dependences for the cross-sections and the analyzing powers of the $(\vec{d},p)$
reaction are in good agreement with corresponding experimental data.
\end{abstract}


\maketitle

\section{Introduction}
\label{secI}
Deuteron-induced nuclear reactions are widely used for studying the spectroscopic properties
and the structure of nuclei. The importance of such reactions in experimental nuclear physics is
associated with both the simplicity in obtaining monochromatic deuteron beams with the precisely
calibrated polarization and a large yield of $dA$-reactions in comparison with reactions induced
by other charged particles.

The majority of theoretical works on deuteron stripping reactions were published in 1950s--1970s.
A surge of interest in those reactions, which has been observed in the last decade, was connected
with intensive researches of radioactive nuclei with an excess of protons or neutrons near the
stability valley. In those reactions, besides the emergence of a residual nucleus in the bound
state, the formation of resonant states also becomes probable~\cite{muk11}. This circumstance
makes the nucleon-transfer reactions a unique tool for studying unstable nuclei and astrophysical
reactions of the $(N,\gamma)$ and $(p,\alpha)$ types.

In this work, the inclusive deuteron stripping reaction was considered. Its formalism for the case
of intermediate energies was proposed for the first time by Serber~\cite{ser47} and developed
in works by Akhiezer and Sitenko~\cite{akh57,sit90}. In work~\cite{kov15}, it was shown that the
multiple integral in the expression for the reaction cross-section~\cite{sit90} can be calculated
analytically by expanding the integrands in the basis of Gaussoid functions. The scattering phases
were calculated at that in the framework of Glauber formalism, making use of the double folding
potential~\cite{cha90,shu01} and the tabulated distributions of the target nucleus
density~\cite{vri87}. Therefore, the final result (the cross-section) depended on a single parameter
corresponding to the normalization of the imaginary part of nucleon-nucleus potential. This
approach was used to describe the differential cross-sections of $^{2}$H$(d,n)^{3}$He
reaction~\cite{wil80} at relativistic energies of incident deuterons. The application of the
deuteron wave function with a correct asymptotic at large distances between nucleons was shown
to be important at calculations.

By continuing work~\cite{kov15}, in this work, a more complicated problem was considered;
namely, the calculation of observables for the inclusive polarized-deuteron stripping reaction.
The microscopic description for the distributions of nucleon density and scattering phases
was also introduced. This procedure made it possible to keep the number of fitting parameters
to a minimum and, in such a manner, to obtain a quantitative description of the experiment.
Besides, the model itself and the limits of its applicability to the kinematics of the reaction
concerned were verified. Notice that, in this approach, the reaction density matrix is a
five-fold integral only formally, because the profile functions, which the density matrix
depends on, are also expressed in terms of multiple integrals. Therefore, generally speaking,
we have rather a complicated computational problem.

Nevertheless, as will be shown below, the final formulas for the cross-section and analyzing
power can be reduced to algebraic expressions, namely, multiple sums of elementary functions,
if Gaussoid functions are used as integrands. Notice that a similar trick is applied rather
often: in the variational approach to the description of bound states~\cite{kuk77,var95,gri00},
for the parametrization of nuclear charge densities in the ground state of nucleus~\cite{sic74},
and in scattering problems~\cite{dal85}, which makes it possible to calculate the corresponding
scattering phases and form factors analytically.

The structure of the paper is as follows. Section~\ref{secII} is devoted to the description
of formalism applied while calculating the angular (energy) distributions of cross-sections
and polarizations for nucleons that arise in the deuteron stripping reaction.
In Section~\ref{secIII}, the results of numerical calculations for the differential cross-sections
and analyzing powers are discussed and compared with corresponding experimental data.
Section~\ref{secIV} contains conclusions. The most important auxiliary formulas used to simplify
the formalism description are given in Appendices.

\section{Formalism}
\label{secII}
Light and medium nuclei were selected as targets, because in this case and in the case of
intermediate energies, the Coulomb interaction can be neglected.

Let a proton be a particle that arises in the deuteron stripping reaction.
The angular distribution of protons arising in the reaction $(\vec{d},p)$ is described by
the formula~\cite{sat83}
\begin{equation}
d\sigma/d\Omega=(1+3\mathbf{P}_{d}\mathbf{P})(d\sigma/d\Omega)_{0}\,,
\label{eq1}
\end{equation}
where $\mathbf{P}_{d}$ is the polarization of incident deuterons, and $\mathbf{P}$
the polarization of protons arising in the non-polarized deuteron stripping reaction.
The cross-section for the latter reaction equals $(d\sigma/d\Omega)_{0}$.
The quantity $\mathbf{P}$ is defined in terms of the density matrix $\rho$
as follows~\cite{sit12}:
\begin{equation}
\mathbf{P}=\frac{\text{Tr\thinspace}({\bm\sigma}\rho)}
{\text{Tr\thinspace}\rho}\,,
\label{eq2}
\end{equation}
where $\bm\sigma$ are Pauli's matrices.

Let the deuteron move in the positive direction of $z$-axis in the Cartesian coordinate system,
so that the $xy$-plane is the impact parameter plane. The proton and neutron of the deuteron
will be designated by subscripts~1 and 2, respectively. In the diffraction approximation,
the general expression for the density matrix of stripping reaction looks like~\cite{sit90}
\begin{equation}
\rho(\mathbf{k}_{1})=\int d\mathbf{b}_{2}(1-|1-\Omega_{2}|^{\,2}
)a_{\mathbf{k}_{1}}(\mathbf{r}_{2})a_{\mathbf{k}_{1}}^{\dagger}(\mathbf{r}_{2}),
\label{eq3}
\end{equation}
where $\mathbf{b}_{2}$ is the neutron impact parameter vector, $\Omega_{2}$ the corresponding
profile function; the quantity
\begin{equation}
a_{\mathbf{k}_{1}}(\mathbf{r}_{2})=\int d\mathbf{r}_{1}
\exp(-i\mathbf{k}_{1}\mathbf{r}_{1})(1-\Omega_{1})\varphi_{0}(\mathbf{r}),\quad
\mathbf{r}=\mathbf{r}_{1}-\mathbf{r}_{2},
\label{eq4}
\end{equation}
is the probability amplitude that the proton has the momentum $\mathbf{k}_{1}$ and the neutron
is located at the point $\mathbf{r}_{2}$, and {$\varphi_{0}(\mathbf{r})$} is the deuteron
wave function. The neutron profile function $\Omega_{2}$ in (\ref{eq3}) contains only the
radial part, i.e. $\Omega_{2}(b_{2})=\omega_{2}(b_{2})$, whereas the proton one,
${\Omega}_{1}$, depends also on the spin~\cite{sit58}:

\begin{equation}
{\Omega}_{1}(\mathbf{r}_{1})=\omega_{1}(b_{1})\{1+i\gamma_{1}
\exp(i\delta_{1}){\bm\sigma}((\mathbf{k}/2)\times\partial/\partial\mathbf{r}_{1})\},
\label{eq5}
\end{equation}
where $b_{1}$, $\gamma_{1}$, and $\delta_{1}$ are the impact parameter, the constant of
spin-orbit interaction between the proton and the nucleus, and the phase shift, respectively;
and $\mathbf{k}$ is the vector of the incident deuteron momentum.

If the integrand in (\ref{eq3}) includes Gaussian functions, the density matrix can be calculated
explicitly. Without loss of generality, let us expand the functions $\omega_{i}(b_{i})$ and
$\varphi_{0}(\mathbf{r})$ in series of Gaussoid basis functions,
\begin{equation}
\omega_{i}(b_{i})=\sum\limits_{j=1}^{N}{\alpha_{ij}}\exp(-b_{i}^{2}/\beta_{ij}),
\label{eq6}
\end{equation}
\vspace{-5mm}
\begin{equation}
\varphi_{0}(\mathbf{r})=\sum\limits_{j=1}^{N}{c_{j}}
\exp(-d_{j}|\mathbf{r}_{1}-\mathbf{r}_{2}|^{2}),
\label{eq7}
\end{equation}
where $\beta_{ij}=R_{rms}^{2}/j$, and $R_{rms}$ is the root-mean-square radius of target nucleus.

Analytical integration in (\ref{eq3}) with functions (\ref{eq6}) and (\ref{eq7}), and the calculation
of the traces of corresponding matrices in the numerator and denominator of formula (\ref{eq2})
bring us to the expressions $\mathrm{Tr\,}({\bm\sigma}\rho(\mathbf{k}_{1}))=
\mathbf{G}(\mathbf{k}_{1})$ and $\mathrm{Tr}\,\rho(\mathbf{k}_{1})=H(\mathbf{k}_{1})$
(see Appendix~\ref{appA}). Therefore, the proton polarization is determined by the formula
\begin{equation}
\mathbf{P}(\mathbf{k}_{1})=\frac{G(\mathbf{k}_{1})}
{H(\mathbf{k}_{1})}({\bm{\kappa}}_{1}\!\times\!{\mathbf{k}_{1}})\,,
\label{eq8}
\end{equation}
where ${\bm{\kappa}}_{1}$ is the transverse component of the momentum
$\mathbf{k}_{1}=\left\{{\bm{\kappa}}_{1},(\mathbf{k}/k)k_{1z}\right\}$. The magnitudes
$\kappa_{1}$ and $k_{1z}$ of the $\mathbf{k}_{1}$-vector components are related to the
proton energy $T_{1}$ and the proton emission angle $\Theta_{1}$ in the laboratory frame
by the relationships~\cite{sit90}
\begin{equation}
\kappa_{1}=(k/2+k_{1z})\tan\Theta_{1},
\label{eq9}
\end{equation}
\begin{equation}
k_{1z}=\sqrt{m/T}(T_{1}-T/2),
\label{eq10}
\end{equation}
where $m$ is the nucleon mass, and $T$ the initial energy of deuteron.

The cross-section of non-polarized-deuteron stripping reaction, after which the wave
vector of emitted proton falls within the interval $d\mathbf{k}_{1}$, is determined by
the denominator in (\ref{eq8}):
\begin{equation}
d\sigma_{0}=H(\mathbf{k}_{1})\frac{d\mathbf{k}_{1}}{(2\pi)^{3}}.
\label{eq11}
\end{equation}
In order to find the dependences of polarization (\ref{eq8}) and cross-section (\ref{eq11})
on the proton emission angle, the expression obtained for $G(\mathbf{k}_{1})$ and
$H(\mathbf{k}_{1})$ have to be integrated over the $z$-component of vector $\mathbf{k}_{1}$~\cite{sit90}.
By expressing the components of $d\mathbf{k}_{1}$ in the cylindrical coordinate system and
taking into account (\ref{eq9}), we obtain
\begin{equation}
P(\Theta_{1})=\frac{\widetilde{G}(\kappa_{1})}{\widetilde{H}(\kappa_{1})}\,,\quad
\sigma_{0}(\Theta_{1})=\frac{\widetilde{H}(\kappa_{1})}
{(2\pi)^{3}\cos^{3}\Theta_{1}}\,,
\label{eq12}
\end{equation}
where
\begin{equation}
\widetilde{G}(\kappa_{1})=\int_{-\infty}^{\infty}(k/2+k_{1z})^{2}G(\kappa_{1},k_{1z})dk_{1z}\,,
\label{eq13}
\end{equation}
\vspace{-3mm}
\begin{equation}
\widetilde{H}(\kappa_{1})=\int_{-\infty}^{\infty}(k/2+k_{1z})^{2}H(\kappa_{1},k_{1z})dk_{1z}\,.
\label{eq14}
\end{equation}
Cross-section (\ref{eq12}) and the angle $\Theta_{1}$ were transformed into their counterparts
in the center-of-mass system making use of kinematic relations from work~\cite{bal63}.

\section{Calculation results and their discussion}
\label{secIII}
The formalism described in the previous section was applied to analyze experimental data (the
cross-section and the analyzing power) obtained for $(\vec{d},p)$ stripping reactions on the
$^{24}\text{Mg}$ and $^{40}\text{Ca}$ nuclei at the deuteron energy $T=56$~MeV~\cite{hat84,uoz94}.

When expanding the wave function $\varphi_{0}(\mathbf{r})$ in series (\ref{eq7}), tabulated data
for the S-component of deuteron wave function obtained with the help of realistic $NN$-potential
Nijm~I~\cite{sto94} were used.

The profile functions $\omega_{i}(b_{i})$, which were expanded in series (\ref{eq6}), were first
calculated in the eikonal approximation (see Appendix~B) making use of the tabulated nuclear
density distributions for the $^{24}\text{Mg}$ and $^{40}\text{Ca}$ nuclei taken from
work~\cite{vri87}. The number $N$ of terms in expansions (\ref{eq6}) and (\ref{eq7}) was taken to
equal 12.

The values of spin-orbit parameters in (\ref{eq5}) were determined from the relations~\cite{sit90}
\begin{equation}
\delta_{1}=\arctan(W_{0}/V_{0}),\quad
\gamma_{1}=V_{\text{so}}r_{\text{so}}^{2}/(2\sqrt{V_{0}^{2}+W_{0}^{2}}),
\label{eq15}
\end{equation}
where $V_{0}$, $W_{0}$, $V_{\text{so}}$, and $r_{\text{so}}$ are the relevant parameters for
the central and spin-orbit parts of the nucleon-nucleus optical potential. From the data of
work~\cite{bor69,fab80}, it follows that, for the proton energy ${T_{1}=T/2=28}$~MeV,
the parameters $\delta_{1}$ and $\gamma_{1}$ are equal to 0.12 and 0.07, respectively,
for the $^{24}$Mg target nucleus, and to 0.13 and 0.08, respectively, for $^{40}$Ca.

The coefficients in expansions (\ref{eq6}) and (\ref{eq7}) together with quantities (\ref{eq15})
were used to calculate the angular distribution of protons,
\begin{equation}
\sigma(\theta)=[1+3P_{d}P(\theta)]\sigma_{0}(\theta),
\label{eq16}
\end{equation}
where $P(\theta)$ and $\sigma_{0}(\theta)$ are the polarization and cross-section (\ref{eq12})
in the center-of-mass system. The polarization of deuteron beam, $P_{d}$, amounted
to about 0.5~\cite{hat84} or 0.52~\cite{uoz94}.

In Fig.~1, cross-sections (\ref{eq16}) and analyzing powers {${A}_{y}(\theta)=2P(\theta)$}~\cite{yul68}
calculated for polarized-deuteron stripping reactions on the $^{24}\text{Mg}$ and $^{40}\text{Ca }$ nuclei
at $T=56$~MeV are depicted by solid curves. The dashed curves were obtained for the model function
\begin{equation}
\varphi_{0}(\mathbf{r})=(2\xi/\pi)^{3/4}
\exp(-\xi|\mathbf{r}_{1}-\mathbf{r}_{2}|^{2}).
\label{eq17}
\end{equation}
The value of the parameter $\xi$ was selected to equal $\xi=0.049$~fm$^{-2}$, at which formula
(\ref{eq17}) reproduced the experimental mean-square radius of deuteron~\cite{mus92}. The dash-dotted
curves correspond to the results of calculations carried out in works~\cite{hat84,uoz94} for the
quantities $\sigma(\theta)$ and $A_{y}(\theta)$ in the framework of the adiabatic distorted
wave approximation~\cite{jsp70}.

\begin{figure}[!h]
\begin{center}
\includegraphics[width=8.6 cm]{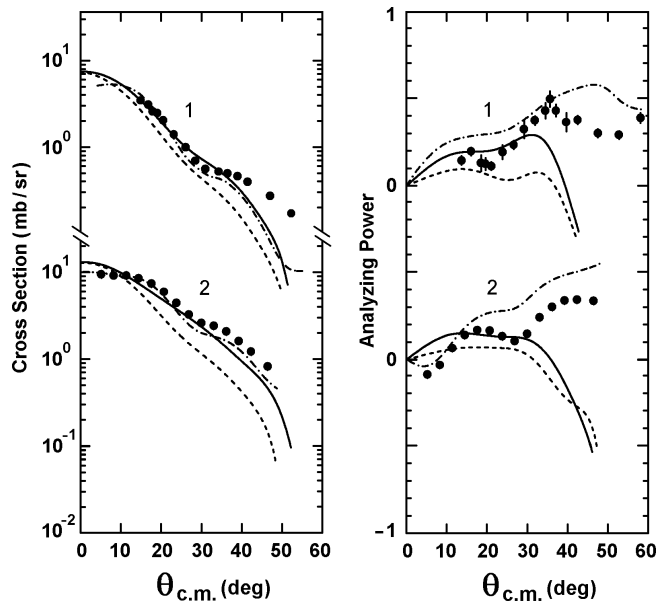}
\end{center}
\caption{Angular dependences of cross-sections (left panel) and analyzing powers (right panel)
for $(\vec{d},p)$ stripping reactions on the $^{24}\text{Mg}$~(1) and $^{40}\text{Ca}$~(2)
nuclei at a deuteron energy of 56~MeV. See further explanations in the text. Experimental data
were taken from works \cite{hat84,uoz94}.
\label{fig1}}
\end{figure}

The experimental data were fitted using a single fitting parameter $N_{W}$, the normalization
parameter for the imaginary part of double folding potential (see Appendix~\ref{appB}).
Its values amounted to 0.24 for the $^{24}$Mg target nucleus and 0.19 for $^{40}$Ca.
By varying the parameter $N_{W}$, the both dependences, $\sigma(\theta)$ and $A_{y}(\theta)$,
were fitted simultaneously to the relevant experimental data. The parameter $N_{W}$ was found
to affect the shape of the curve $A_{y}(\theta)$, but not the slope of the linear section in
the dependence $\sigma(\theta)$.

Direct nuclear reactions, including the stripping one, are surface reactions~\cite{sat83}.
Therefore, the calculated observable quantities are expected to be sensitive to the asymptotics
of the wave functions of interacting particles, which is confirmed by specific calculations.
From Fig.~1, one can see that deuteron model function (\ref{eq17}), which has an incorrect
asymptotic at large nucleon-nucleon distances, does not allow the cross-sections and analyzing
powers to be described adequately (the dashed curves).

Notice that the deuteron stripping problem considered above was solved here exactly, making no
additional simplifications and restrictions, which could affect the result of numerical calculation.
From the analysis of the behavior of solid curves in Fig.~1, it follows that the calculation
results satisfactorily describe the corresponding experimental data for the angles
{$\theta<(30\div35)^{\circ}$}. Therefore, in the case of the incident deuteron energy $T=56$~MeV
and the $^{24}\text{Mg}$ and $^{40}\text{Ca target }$nuclei, the applicability region of
diffraction approximation for the description of $(\vec{d},p)$ reaction is confined to the
indicated interval of proton emission angles.

\section{Conclusions}
\label{secIV}

The main result of this work includes general expressions (Appendix~\ref{appA}) that allow the cross-sections
and the analyzing powers of the inclusive polarized-deuteron stripping reaction to be calculated.
The formulas were obtained in the diffraction approximation by analytically integrating the
expression for the corresponding density matrix. This approach can also be applied in other
similar problems with multiple integrals if the integrand allows an expansion in a series of
Gaussoid basis functions.

With the help of tabulated data for the densities of target nuclei and making use of a realistic
deuteron wave function with a correct asymptotic at large nucleon-nucleon distances, the observed
angular dependences of the cross-sections and analyzing powers for the $(\vec{d},p)$ reaction were
described in the case of $^{24}\text{Mg}$ and $^{40}\text{Ca target}$ nuclei and a deuteron energy
of 56~MeV. A single fitting parameter was used at that, namely, the normalization parameter for
the imaginary part of the high-energy double folding potential.

It should be noticed that the general formulas obtained in this work can be used to describe the
cross-sections and polarizations not only for the inclusive processes of deuteron stripping,
but also deuteron, as well as light- and heavy-ion, breakup~\cite{uts85}. Both the angular and energy
distributions of indicated quantities can be calculated. In the latter case, expressions for
$G$ and $H$ in (\ref{eq8}) and (\ref{eq11}) should be integrated over the normal components of
the emitted particle momentum.

\appendix
\section{Traces of the density matrix $\rho$ and the products {\protect\normalsize {${\bm\sigma}\rho$}}}
\label{appA}
In this Appendix, the calculation results are presented for the traces of the matrices in the
numerator and denominator of expression (\ref{eq2}) after analytical integration in (\ref{eq3})
with functions (\ref{eq6}) and (\ref{eq7}).

The numerator in (\ref{eq2}) looks like
\begin{equation}
\text{Tr}\,({\bm\sigma}\rho(\mathbf{k}_{1}))=\mathbf{G}(\mathbf{k}_{1})=
G(\mathbf{k}_{1})({\bm{\kappa}}_{1}\!\times\!{\mathbf{k}_{1}})\,.
\label{A1}
\end{equation}
Here, {${\bm{\kappa}}_{1}$} is the transverse component of emitted particle momentum
{$\mathbf{k}_{1}=\left\{{\bm{\kappa}}_{1},\mathbf{k}_{1z}\right\}$}, and the quantity
$G(\mathbf{k}_{1})$ is defined as follows:
\begin{equation}
G(\mathbf{k}_{1})=\gamma_{1}\sin\delta_{1}\sum\limits_{p=1}^{N}
\sum\limits_{q=1}^{N}{c_{p}}{c_{q}}Z(\lambda,\kappa_{1},k_{1z}),
\label{A2}
\end{equation}
where
\begin{equation}
Z(\lambda,\kappa_{1},k_{1z})=2\lambda t(\lambda,k_{1z})(4z^{(1)}
(\lambda,\kappa_{1})-z^{(2)}(\lambda,\kappa_{1})),
\label{A3}
\end{equation}
\begin{equation}
z^{(1)}(\lambda,\kappa_{1})= z_{11}(\lambda,\kappa_{1})-z_{12}
(\lambda,\kappa_{1}),
\label{A4}
\end{equation}
\begin{equation}
z^{(2)}(\lambda,\kappa_{1})= 4z_{21}(\lambda,\kappa_{1})+z_{22}
(\lambda,\kappa_{1}).
\label{A5}
\end{equation}

The denominator in (\ref{eq2}) is determined by the formula
\begin{equation}
\text{Tr}\,\rho(\mathbf{k}_{1})=H(\mathbf{k}_{1})=
H_{0}(\mathbf{k}_{1})+\gamma_{1}^{2}H_{\text{so}}(\mathbf{k}_{1})\,,
\label{A6}
\end{equation}
where
\begin{equation}
H_{0}(\mathbf{k}_{1})=\sum\limits_{p=1}^{N}\sum\limits_{q=1}^{N}{c_{p}}{c_{q}}
Y(\lambda,\kappa_{1},k_{1z}),
\label{A7}
\end{equation}

\begin{equation}
Y(\lambda,\kappa_{1},k_{1z})=t(\lambda,k_{1z})
(y^{(1)}(\lambda,\kappa_{1})-y^{(2)}(\lambda,\kappa_{1})),
\label{A8}
\end{equation}

\begin{equation}
y^{(1)}(\lambda,\kappa_{1})=4(y_{11}(\lambda,\kappa_{1})-
2y_{12}(\lambda,\kappa_{1})+y_{13}(\lambda,\kappa_{1})),
\label{A9}
\end{equation}

\begin{equation}
y^{(2)}(\lambda,\kappa_{1})=y_{21}(\lambda,\kappa_{1})-
4y_{22}(\lambda,\kappa_{1})+y_{23}(\lambda,\kappa_{1});
\label{A10}
\end{equation}
\begin{equation}
H_{\text{so}}(\mathbf{k}_{1})= \sum\limits_{p=1}^{N}
\sum\limits_{q=1}^{N}{c_{p}}{c_{q}}W(\lambda,\kappa_{1},k_{1z}),
\label{A11}
\end{equation}
\begin{eqnarray}
W(\lambda,\kappa_{1},k_{1z})=\,\,&&t(\lambda,k_{1z})(4w^{(1)}
(\lambda,\kappa_{1},k_{1z})\nonumber\\
&&-
w^{(2)}(\lambda,\kappa_{1},k_{1z})),
\label{A12}
\end{eqnarray}
\begin{equation}
w^{(1)}(\lambda,\kappa_{1},k_{1z})= w_{11}(\lambda,\kappa_{1},k_{1z})+
2w_{12}(\lambda,\kappa_{1},k_{1z}),
\label{A13}
\end{equation}
\begin{equation}
w^{(2)}(\lambda,\kappa_{1},k_{1z})= w_{21}(\lambda,\kappa_{1},k_{1z})+
w_{22}(\lambda,\kappa_{1},k_{1z}).
\label{A14}
\end{equation}

\vspace{5mm}
The values in right sides of (\ref{A4}), (\ref{A5}), (\ref{A9}), (\ref{A10}), (\ref{A13}),
and (\ref{A14}) are defined as follows:

\begin{widetext}
\begin{equation}
z_{11}(\lambda,\kappa_{1})=\sum\limits_{i=1}^{N}\sum\limits_{j=1}^{N}
\frac{\alpha_{1i}\beta_{1i}\,\alpha_{2j}\beta_{2j}}
{(\lambda+\beta_{1i}+\beta_{2j})^{2}}
\exp\Bigl(-\frac{\lambda+2\beta_{1i}+2\beta_{2j}}
{\lambda+\beta_{1i}+\beta_{2j}}\,\frac{\lambda\kappa_{1}^{2}}{4}\Bigr),
\label{A15}
\end{equation}
\begin{equation}
z_{12}(\lambda,\kappa_{1})=\sum\limits_{i=1}^{N}\sum\limits_{j=1}^{N}\sum\limits_{l=1}^{N}
\frac{\alpha_{1i}\beta_{1i}\,\alpha_{1j}\beta_{1j}\,\alpha_{2l}\,\beta_{2l}}
{(\lambda+\beta_{1ij})^{2}(\lambda+\beta_{1ij}+2\beta_{2l})}
\exp\Bigl(-\frac{\beta_{1ij}}{\lambda+\beta_{1ij}}\,
\frac{\lambda\kappa_{1}^{2}}{2}\Bigr),
\label{A16}
\end{equation}
\begin{equation}
z_{21}(\lambda,\kappa_{1})=\sum\limits_{i=1}^{N}\sum\limits_{j=1}^{N}\sum\limits_{l=1}^{N}
\frac{\alpha_{1i}\beta_{1i}\,\alpha_{2j}\,\alpha_{2l}\beta_{2jl}}
{(2\lambda+2\beta_{1i}+\beta_{2jl})^{2}}
\exp\Bigl(-\frac{\lambda+2\beta_{1i}+\beta_{2jl}}
{2\lambda+2\beta_{1i}+\beta_{2jl}}\,\frac{\lambda\kappa_{1}^{2}}{2}\Bigr),
\label{A17}
\end{equation}
\begin{equation}
z_{22}(\lambda,\kappa_{1})=\sum\limits_{i=1}^{N}\sum\limits_{j=1}^{N}
\sum\limits_{l=1}^{N}\sum\limits_{n=1}^{N}
\frac{\alpha_{1i}\beta_{1i}\,\alpha_{1j}\beta_{1j}\,\alpha_{2l}\alpha_{2n}\beta_{2ln}}
{(\lambda+\beta_{1ij})^{2}(\lambda+\beta_{1ij}+\beta_{2ln})}
\exp\Bigl(-\frac{\beta_{1ij}}{\lambda+\beta_{1ij}}\,
\frac{\lambda\kappa_{1}^{2}}{2}\Bigr);
\label{A18}
\end{equation}
\begin{equation}
y_{11}(\lambda,\kappa_{1})=\exp\Bigl(-\frac{\lambda\kappa^{2}_{1}}{2}\Bigr)
\sum\limits_{i=1}^{N}\alpha_{2i}\beta_{2i},
\label{A19}
\end{equation}
\begin{equation}
y_{12}(\lambda,\kappa_{1})=\sum\limits_{i=1}^{N}\sum\limits_{j=1}^{N}
\frac{\alpha_{1i}\beta_{1i}\,\alpha_{2j}\beta_{2j}}{\lambda+\beta_{1i}+\beta_{2j}}
\exp\Bigl(-\frac{\lambda+2\beta_{1i}+2\beta_{2j}}
{\lambda+\beta_{1i}+\beta_{2j}}\,\frac{\lambda\kappa^{2}_{1}}{4}\Bigr),
\label{A20}
\end{equation}
\begin{equation}
y_{13}(\lambda,\kappa_{1})=\sum\limits_{i=1}^{N}\sum\limits_{j=1}^{N}\sum\limits_{l=1}^{N}
\frac{\alpha_{1i}\beta_{1i}\,\alpha_{1j}\beta_{1j}\,\alpha_{2l}\beta_{2l}}
{(\lambda+\beta_{1ij})(\lambda+\beta_{1ij}+2\beta_{2l})}
\exp\Bigl(-\frac{\beta_{1ij}}{\lambda+\beta_{1ij}}\,
\frac{\lambda\kappa^{2}_{1}}{2}\Bigr),
\label{A21}
\end{equation}
\begin{equation}
y_{21}(\lambda,\kappa_{1})=\exp\Bigl(-\frac{\lambda\kappa^{2}_{1}}{2}\Bigr)
\sum\limits_{i=1}^{N}\sum\limits_{j=1}^{N}\alpha_{2i}\beta_{2i}\,\beta_{2ij},
\label{A22}
\end{equation}
\begin{equation}
y_{22}(\lambda,\kappa_{1})=\sum\limits_{i=1}^{N}\sum\limits_{j=1}^{N}\sum\limits_{l=1}^{N}
\frac{\alpha_{1i}\beta_{1i}\,\alpha_{2j}\beta_{2j}\,\beta_{2jl}}
{2\lambda+2\beta_{1i}+\beta_{2jl}}
\exp\Bigl(-\frac{\lambda+2\beta_{1i}+\beta_{2jl}}{2\lambda+2\beta_{1i}+\beta_{2jl}}\,
\frac{\lambda\kappa^{2}_{1}}{2}\Bigr),
\label{A23}
\end{equation}
\begin{equation}
y_{23}(\lambda,\kappa_{1})=\sum\limits_{i=1}^{N}\sum\limits_{j=1}^{N}
\sum\limits_{l=1}^{N}\sum\limits_{n=1}^{N}
\frac{a_{1i}\beta_{1i}\,a_{1j}\beta_{1j}\,a_{2l}\beta_{2l}\,\beta_{2ln}}
{(\lambda+\beta_{1ij})(\lambda+\beta_{1ij}+\beta_{2ln})}
\exp\Bigl(-\frac{\beta_{1ij}}{\lambda+\beta_{1ij}}\,\frac{\lambda\kappa^{2}_{1}}{2}\Bigr);
\label{A24}
\end{equation}
\begin{equation}
w_{11}(\lambda,\kappa_{1},k_{1z})=
\lambda^{2}\kappa^{2}_{1}k^{2}_{1z}\sum\limits_{i=1}^{N}\sum\limits_{j=1}^{N}\sum\limits_{l=1}^{N}
\frac{\alpha_{1i}\beta_{1i}\,\alpha_{1j}\beta_{1j}\,\alpha_{2l}\beta_{2l}}
{(\lambda+\beta_{1ij})^{3}(\lambda+\beta_{1ij}+2\beta_{2l})}
\exp\Bigl(-\frac{\beta_{1ij}}{\lambda+\beta_{1ij}}\,
\frac{\lambda\kappa_{1}^{2}}{2}\Bigr),
\label{A25}
\end{equation}
\begin{equation}
w_{12}(\lambda,\kappa_{1},k_{1z})=
(2k_{1z}^{2}+\kappa_{1}^{2})\sum\limits_{i=1}^{N}\sum\limits_{j=1}^{N}\sum\limits_{l=1}^{N}
\frac{\alpha_{1i}\beta_{1i}\,\alpha_{1j}\beta_{1j}\,\alpha_{2l}\,\beta_{2l}^{2}}
{(\lambda+\beta_{1ij})^{2}(\lambda+\beta_{1ij}+2\beta_{2l})^{2}}
\exp\Bigl(-\frac{\beta_{1ij}}{\lambda+\beta_{1ij}}\,
\frac{\lambda\kappa_{1}^{2}}{2}\Bigr),
\label{A26}
\end{equation}
\begin{equation}
w_{21}(\lambda,\kappa_{1},k_{1z})=\lambda^{2}\kappa^{2}_{1}k^{2}_{1z}
\sum\limits_{i=1}^{N}\sum\limits_{j=1}^{N}\sum\limits_{l=1}^{N}\sum\limits_{n=1}^{N}
\frac{\alpha_{1i}\beta_{1i}\,\alpha_{1j}\beta_{1j}\,\alpha_{2l}\alpha_{2n}\beta_{2ln}}
{(\lambda+\beta_{1ij})^{3}(\lambda+\beta_{1ij}+\beta_{2ln})}
\exp\Bigl(-\frac{\beta_{1ij}}{\lambda+\beta_{1ij}}\,
\frac{\lambda\kappa_{1}^{2}}{2}\Bigr),
\label{A27}
\end{equation}
\begin{equation}
w_{22}(\lambda,\kappa_{1},k_{1z})=(2k_{1z}^{2}+\kappa_{1}^{2})
\sum\limits_{i=1}^{N}\sum\limits_{j=1}^{N}\sum\limits_{l=1}^{N}\sum\limits_{n=1}^{N}
\frac{\alpha_{1i}\beta_{1i}\,\alpha_{1j}\beta_{1j}\,\alpha_{2l}\alpha_{2n}\beta_{2ln}^{2}}
{(\lambda+\beta_{1ij})^{2}(\lambda+\beta_{1ij}+\beta_{2ln})^{2}}
\exp\Bigl(-\frac{\beta_{1ij}}{\lambda+\beta_{1ij}}\,
\frac{\lambda\kappa_{1}^{2}}{2}\Bigr).
\label{A28}
\end{equation}
\end{widetext}

In (\ref{A2}), (\ref{A7}), and (\ref{A11}), $\lambda=2/(d_{p}+d_{q})$. The function
$t(\lambda,k_{1z})$ in (\ref{A3}), (\ref{A8}), and (\ref{A12}) looks like
\begin{equation}
t(\lambda,k_{1z})=\pi^{4}\lambda^{3}\exp\left(-\frac{\lambda k_{1z}^{2}}{2}\right).
\label{A29}
\end{equation}
Besides,
\begin{equation}
\beta_{ijl}=2\beta_{ij}\beta_{il}/(\beta_{ij}+\beta_{il}),\quad
(i\!=\!1,2;\,\,\,j,l\!=\!\overline{1,N}).
\label{A30}
\end{equation}

\section{Calculation of profile functions}
\label{appB}

The radial parts of nucleon-nucleus profile functions were calculated in the eikonal approximation:
\begin{equation}
\omega_{i}(b_{i})=1-\exp[-\phi_{i}(b_{i})],\quad i\!=\!1,2;
\label{B1}
\end{equation}
where
\begin{equation}
\phi_{i}(b_{i})=-\frac{1}{v}\int_{-\infty}^{\infty}dz\,W\left(\sqrt{b_{i}^{2}+z^{2}}\right)
\label{B2}
\end{equation}
is the scattering phase, $v$ the velocity of incident nucleon, and $W(r)$ the imaginary part of
nucleon-nucleus potential.

In the framework of the double folding model, the eikonal phase can be calculated using the method
described in work~\cite{cha90}. Let the distribution of nuclear density in the nucleon,
$\rho_{i}(r)$, and the amplitude of $NN$-interaction at the impact parameter plane, $f(b)$,
be defined by Gaussian functions:
\begin{equation}
\rho_{i}(r)=\rho_{i}(0)\exp(-r^{2}/a_{i}^{2}),
\label{B3}
\end{equation}
\begin{equation}
f(b)=(\pi r_{0}^{2})^{-1}\exp(-b^{2}/r_{0}^{2}),
\label{B4}
\end{equation}
where $\rho_{i}(0)=(a_{i}\sqrt{\pi})^{-3}$, $a_{i}^{2}=r_{0}^{2}=2r_{NN}^{2}/3$,
and $r_{NN}^{2}\cong~0.65~\mathrm{fm}^{2}$ is the mean-square radius of $NN$-interaction. If the
density distribution (tabulated~\cite{vri87} or model) in the target nucleus can be expanded in
a series of Gaussoid basis functions,
\begin{equation}
\rho_{T}(r)=\sum_{j=1}^{N}\rho_{Tj}\exp(-r^{2}/a_{Tj}^{2}),\quad
a_{Tj}^{2}=R_{rms}^{2}/j\,,
\label{B5}
\end{equation}
where $R_{rms}$ is the root-mean-square radius of the nucleus, the formula for the eikonal phase
from work~\cite{cha90} can be generalized~\cite{kov15} to the expression
\begin{equation}
\phi_{i}(b_{i})=N_{W}\sqrt{\pi}\,\bar{\sigma}_{NN}\sum_{j=1}^{N}
\frac{\rho_{Tj}\,a_{Tj}^{3}}{a_{Tj}^{2}+2r_{0}^{2}}
\exp\left(-\frac{b_{i}^{2}}{a_{Tj}^{2}+2r_{0}^{2}}\right),
\label{B6}
\end{equation}
where $N_{W}$ is the normalization factor for the imaginary part of the double folding potential,
and $\bar{\sigma}_{NN}$ is the isotopically averaged cross-section of nucleon-nucleon
interaction~\cite{shu01}.

Formula (\ref{B6}) was used directly while calculating profile functions (\ref{B1}). Afterwards,
they were expanded in the Gaussoid basis (see (\ref{eq6})).


\end{document}